\documentclass[a4paper]{jpconf}
\usepackage{graphicx}
\graphicspath{{pictures/}}
\usepackage{epstopdf}
\usepackage{citesort}

\begin{document}
\title{The simplest model for non-congruent fluid-fluid phase transition in Coulomb system}

\author{N E Stroev$^1$, I L Iosilevskiy$^2$}

\address{Joint Institute for High Temperatures of the Russian Academy of Sciences, Izhorskaya 13, Moscow 125412, Russia}
\address{Moscow Institute of Physics and Technology, Institutskiy Pereulok 9, Dolgoprudny 141700, Russia}

\ead{$^1$ nikita.stroev@phystech.edu,$^2$iosilevskiy@gmail.com}

\begin{abstract}
The simplest model for non-congruent phase transition of gas-liquid type was developed in frames of modified model with no associations of a binary ionic mixture (BIM) on a homogeneous compressible ideal background (or non-ideal) electron gas /BIM($\sim$)/. The analytical approximation for equation of state equation of state of Potekhin and Chabrier of fully ionized electron-ionic plasma was used for description of the ion-ion correlations (Coulomb non-ideality) in combination with ``linear mixture'' (LM) approximation. Phase equilibrium for the charged species was calculated according to the Gibbs-Guggenheim conditions. The presently considered BIM($\sim$) model allows to calculate full set of parameters for phase boundaries of non-congruent variant of phase equilibrium and to study all features for this non-congruent phase transition realization in Coulomb system in comparison with the simpler (standard) forced-congruent evaporation mode. In particular, in BIM($\sim$) there were reproduced two-dimensional remarkable (``banana-like'') structure of two-phase region $P-T$ diagram and the characteristic non-monotonic shape of caloric phase enthalpy-temperature diagram, similar to the non-congruent evaporation of reactive plasma products in high-temperature heating with the uranium-oxygen system. The parameters of critical points (CP) line were calculated on the entire range of proportions of ions $0<X<1$, including two reference values, when CP coincides with a point of extreme temperature and extreme pressure, $X_{T}$ and $X_{P}$. Finally, it is clearly demonstrated the low-temperature property of non-congruent gas--liquid transition---``distillation'', which is weak in chemically reactive plasmas.
\end{abstract}

\section {The Non-congruence  and it's specification in Coulomb systems}

The aim of this work is to study the non-congruent effects on the example of the simplest Coulomb model of ion mixture. The non-congruence is the ability of the system to reach equilibrium in the process of searching for the minimum of the thermodynamic potential by phase separation into subsystems with differing stoichiometry, while maintaining a given stoichiometry of the whole system. Formally, this option is equivalent to the appearance of the additional degree of freedom of a two-phase equilibrium system \cite{1}. The non-congruent phase transition type is well known in some communities, but is almost unknown in others. Because of the Coulomb interactions in the system, it has its own specific features, which are mainly affect the equilibrium conditions (Gibbs-Guggenheim \cite{1,2},  common conditions for chemical, ionization and phase equilibrium) and interphase electrostatic potential (Galvani potential).
	The main general rule for every phase transition in system, consisting of two or more chemical elements, is that such transition has to be non-congruent.
There is no contradiction in typical H$_{2}$O, NH$_{3}$, CH$_{4}$ systems at room temperatures  – due to mono-molecular structure of liquid and gas, in comparison with the same systems  in planetary conditions. Key features of non-congruence are:

-the complementary coexistence of two phase transition scenarios;

 -partial balanced (equal compositions, that are forced to be with a such stoichiometry, or so-called forced-congruent regime (see further)) and full non-congruent equilibrium with thermodynamic different compositions;

-the  increased dimension of phase boundaries in intensive thermodynamic variables ($P-T$, $\mu -T$ etc.) in comparison with the standard one-dimension curves; 

-the expansion of 2D zones of 2-phase coexistence in extensive variables ($P-V$, $T-V$, $X-T$); 

-another properties and another location of critical point; 

-the appearance of several so-called ``end-points" --- points with maximum pressure, temperature or electrochemical potential;

-the appearance of  composition diagrams; 

-the appearance or absence of azeotropic properties.

Some of these properties will be demonstrated in this work and their general description with the Coulomb specification can be found in \cite{1,3}.

\section {Basic Coulomb models}
	One of approaches for the study of plasma thermodynamics and phase transitions is to distinguish two groups of models, which are extreme cases due to the Coulomb interaction --- the systems without phase transitions, but which allow the formation of associations, and systems, that allow phase separation but are without groups or associations of components. The approach of this work is based on the second class of models\cite{1}. The basis of this family is a one-component model of ions (or electrons) in the ``rigid" (incompressible) compensating background of opposite charge (further OCP($\#$)), see \cite{4,5}). For example, such system can be divided into an ideal (or a non-ideal) gas of particles of one type, their correlations and rigid background of particles of another sort, not involved in the identified correlations. Based on this example the hierarchy of models without associations was built. The next step in the evolution of the models is to change the background properties --- namely, the transition to a homogeneous compressible compensating background instead of rigid (denoted by OCP($\sim$)). This leads to the additional phase transitions in system (such as gas-liquid transition with an upper critical point), in addition to a single transition in the OCP($\#$) --- crystallization. Further prototype's  complicating and model range can be generated by sequential addition of the missing system's correlation \cite{1}. Combining the two previous models, taken into account their one-sort correlations, will lead to Double OCP generalization --- superposition of two interacting subsystems only in average. The last step is the ``activation" of correlations between different varieties of particles, the description of this effect can be accomplished in a linear response or in the more complicated non-linear case.

This hierarchy can be used for presentation and numerical calculations of plasma properties, mainly consisting of ions (or particles) of one kind. Describing the mixture requires a prototype model for the one-component system and the mixing rule. Among these rules the most trivial is linear mixing rule (LMR or LM): $$f_{ex}(Z_{1},Z_{2},\Gamma_{e},X_{1})\approx x_{1}f_{ex}(\Gamma_{1},X_{1}=1)+(1-X_{1})f_{ex}(\Gamma_{2},X_{1}=0) ,$$ which was used in this study and is taken from \cite{6}. The ordinary binary ion system (BIM($\#$)) with the same ``rigid" background electrons, which excludes it's (background) thermodynamic properties) is usually based on the OCP($\#$) and LM rule. The following stage is a modified model of BIM($\sim$) with uniformly--compressible compensating background, which is essentially a generalization of DOCP. The same method is used in building the hierarchy of models by including different sort's correlations, described either in the linear approximation, or in more complex way.

\section {Phase transitions in the OCP, DOCP, BIM and the description of the model}
	As mentioned above, the basis for the construction of the Coulomb models (without associations) hierarchy is OCP($\#$), which contains only one phase transition --- crystallization. Characteristic parameter for this transition is the coupling parameter of the plasma, in this case having the value of nearly 175. If you change the background properties, new phase transitions appear --- melting and sublimation. Thus, in the OCP($\sim$)  model and the DOCP, and more complex the standard complete set of phase transitions is presented. Characteristic pattern can be seen in figure 1.

\begin{figure}[h]
\begin{minipage}{17pc}
\includegraphics[width=17pc]{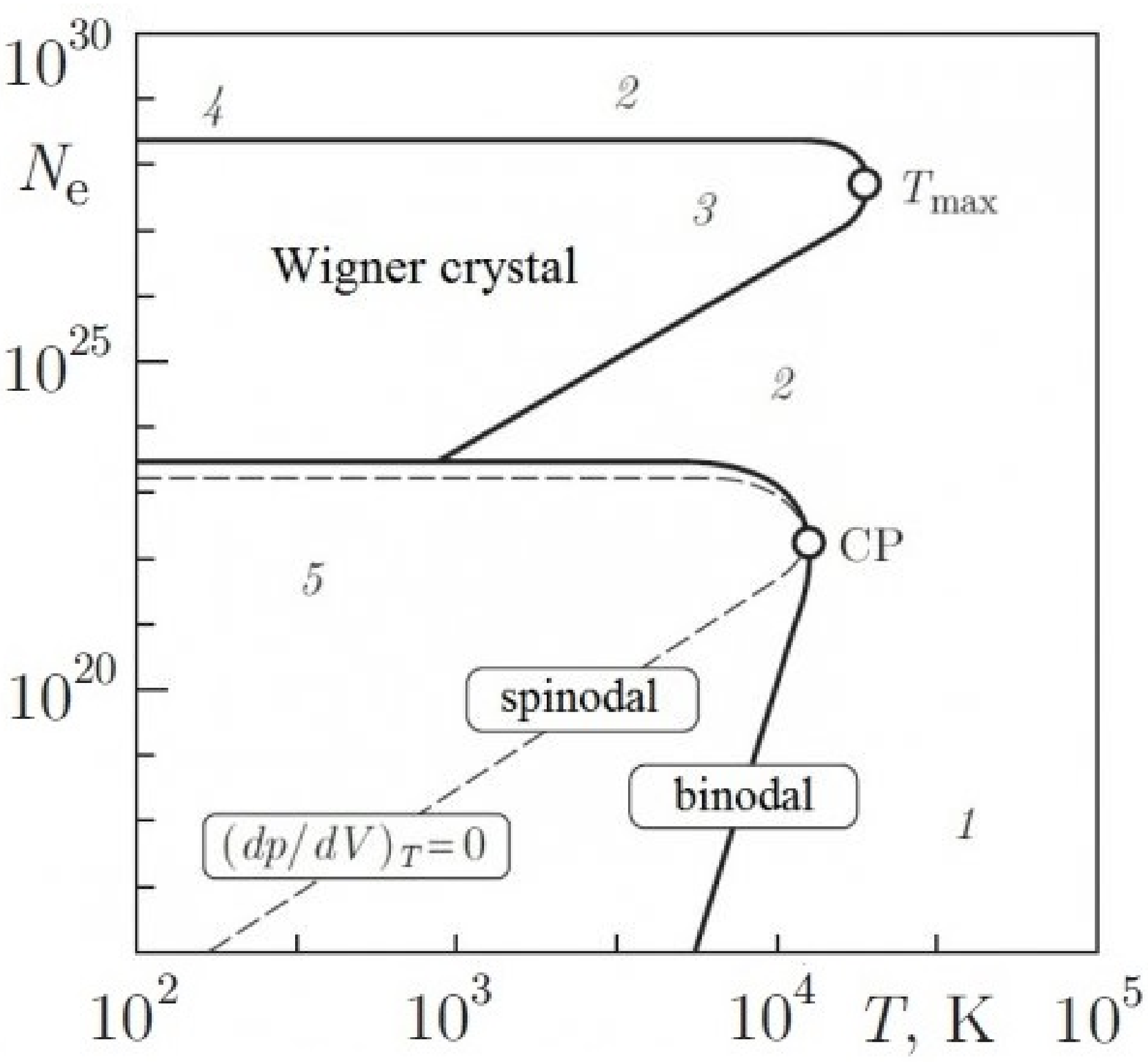}
\caption{\label{label1}Phase diagram for DOCP model of electron-proton plasma. 1 --- gase phase, 2 --- dense phase (liquid type), 3 --- crystal, 4 --- border for cold melting of proton crystal, 5 --- coexisting phases region\cite{7}.}
\end{minipage}\hspace{2pc}%
\begin{minipage}{17pc}
\includegraphics[width=17pc]{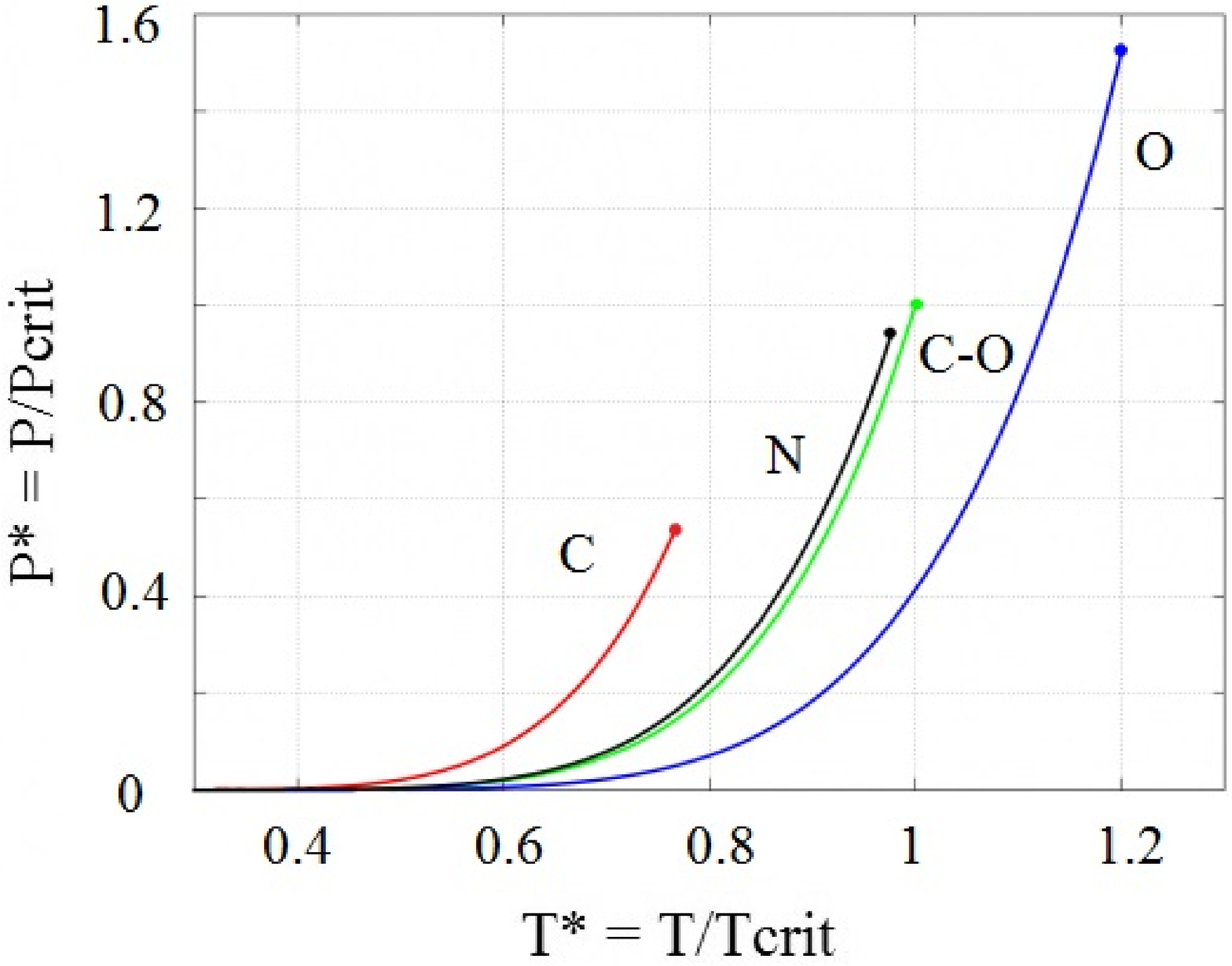}
\caption{\label{label2}P*-T* diagram for carbon, nitrogen, carbon-oxygen mixture and oxygen from left to right with critical points and 1 pseudocritical (C-O).}
\end{minipage} 
\end{figure}

The complete picture of the phase diagrams, which would include standard phase transitions --- melting, sublimation and evaporation with the non-congruence, for mixtures still not been provided (in this paper the attention is given to the  gas-liquid type transition). The behavior of such diagrams at the junction of phase transition types is also uncertain, like the behavior of immiscibility gaps of fluid and solid substances and such exotic transitions as quasicrystalline state transformation and fluid-glass transformation ---  amorphization are also not covered nowadays.

In this paper, a prototype for the construction of a mixture model is an OCP($\sim$), while using the rules of LM to obtain BIM($\sim$) model. The ideal Fermi gas of electrons is taken as the background. Only the ion-ion amendment are taken as correlations\cite{8}. The equilibrium conditions are equality of temperatures, pressures and the Gibbs-Guggenheim conditions\cite{2} --- namely, equality of generalized electrochemical potentials of constituent phases components. Carbon and oxygen were selected for ion components with charged number $Z_{1}=6$ and $Z_{2}=8$ respectively. Electrons do not add a new degree of freedom, as they are connected with the ions by the strict relation of electroneutrality: $n_{e}=Z_{1}n_{i1}+Z_{2}n_{i2}$.
As the main thermodynamic function the Helmholtz free energy was chosen\cite{8}:

$$F(N_{i1},N_{i2},N_{e},V,T)=F_{i1}^{id}+F_{i2}^{id}+F_{e}^{id}+F_{ii}^{ex},$$ where $F_{}^{id}$ is the ideal part (of ions and electrons) and the $F_{}^{ex}$ corresponds to the non-ideality.

For comparison, there are two classes of simplified models, which are also used in this paper: a forced congruent equilibrium regime (FCE regime) and an average ion model (AIM). Typical conditions for the equilibrium of the gas-liquid transition (or fluid-fluid in this case), namely --- an equal phases pressure and equal electrochemical potentials (as described above) in a single component system can be applied to the mixture model, while the potential equality of the phases of one component is replaced with potential equality of the phases of components' complex or cells. In this case, we get the above mentioned Forced congruent equilibrium mode, which corresponds to the model of  ``frozen diffusion", namely the absence of transitions between the phases of the components while maintaining a fixed stoichiometry. For the parameters' values, corresponding to the equilibrium, instead of optionally using electrochemical potentials explicitly, sometimes it's  enough and easier  to apply the provisions of the well-known Maxwell equal areas rule. The second is the average ion model, which meaning is in replacing the Coulomb mixture into a one-component model (single OCP($\sim$) for example), but with components having charge, considering for quite trivial formula $<Z>=Z_{1}X+Z_{2}(1-X) $, where $X$ is the the ratio of carbon ions concentration to the total ions concentration. As shown in this work, models are quite close with each other in the results obtained, especially in a case of weak asymmetry of the charge, but much different from the general non-congruent case\cite{1}.

In this work there is also comparison with other qualitatively different models, where the non-congruence plays a significant role. This is high-temperature non-congruent evaporation of uranium dioxide system in works \cite{1,9,10} and phase transitions in ultra-dense nuclear matter (work \cite{11}).

\section {The FCE and the AIM calculations}

Presented BIM($\sim$) model allows to calculate full set of parameters of phase boundaries of non-congruent regime of equilibrium and to study all properties of such transition realization in Coulomb system (and especially in comparison with FCE and AIM modes, described above). For this model it is usual to expect one-dimensional characteristic curve in the $P-T$ phase diagram (see figure 2). It is essential, that the end point of this curve for mixture is pseudocritical (pseudocritical point --- PCP, or congruent critical point --- CCP further), because of the simplifications, taken in this mode of equilibrium, and is much  different from true critical point, which is considered further.

\begin{figure}[h]
\begin{minipage}{36pc}
\includegraphics[width=36pc]{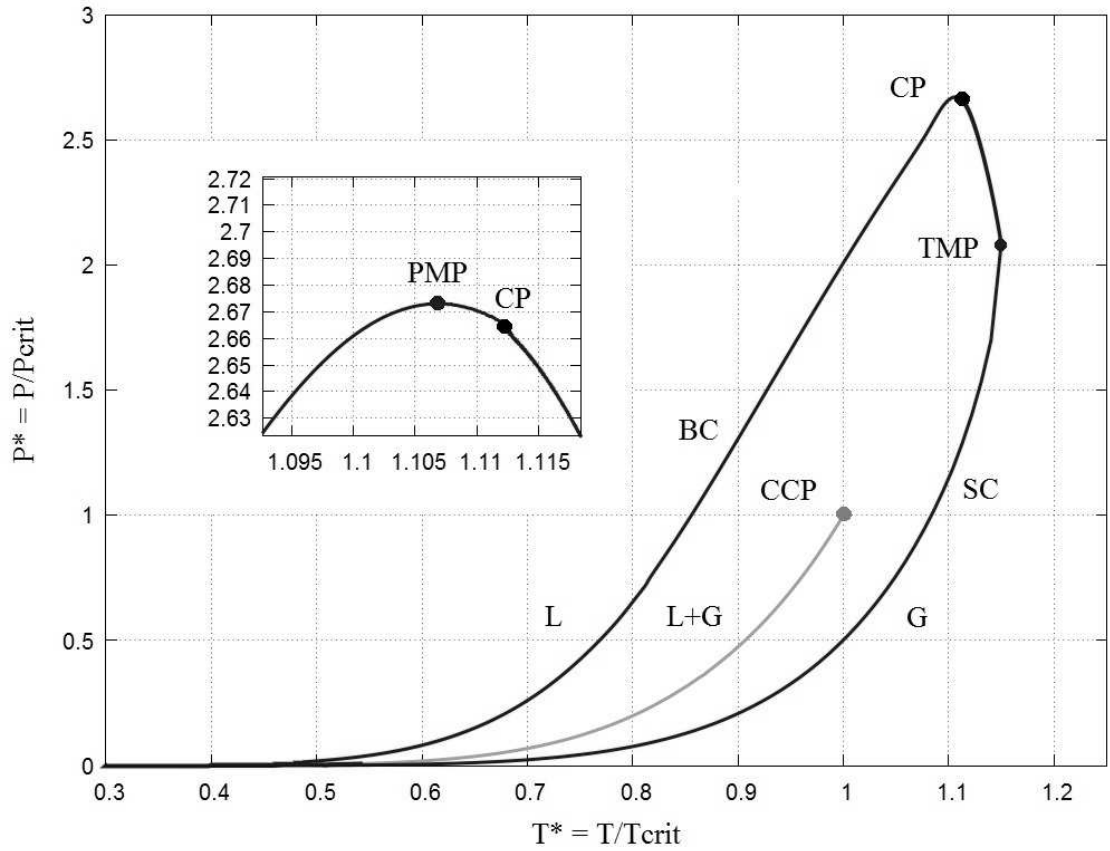}
\caption{\label{label3}P*-T* diagram for fully ionized equimolar C-O plasma.  Grey line – P*(T*) phase boundary in forced-congruent regime (Maxwell double-tangent construction). Black lines --- boundary of two-phase region in non-congruent regime (GG-conditions). CP and CCP --- corresponding critical points (parameters of pseudo-critical point are taken as scaling norm). PMP and TMP --- endpoints of maximal pressure and temperature at P*(T*), BC ---boiling curve, SC-saturation curve, L, G, L+G are for liquid, gas and liquid + gas phases. The difference between PMP and CP is shown in the incut.}
\end{minipage}\hspace{1pc}%
\end{figure}

\begin{figure}[h]
\includegraphics[width=20pc]{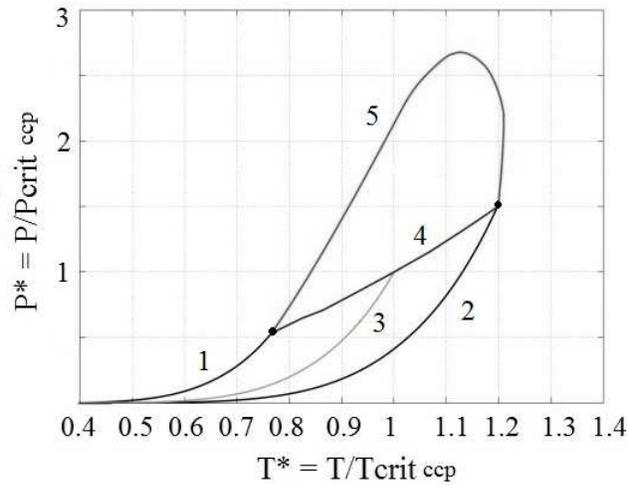}\hspace{2pc}%
\begin{minipage}[b]{16pc}\caption{\label{label4}P*-T* diagram for C-O mixture in two modes of equilibrium and for pure components, and also CP line in two modes. 1 and 2 represent phase boundaries for pure components, 3 --- border for mixture in FCE mode, 4 --- CP line in FCE mode, 5 --- non-congruent CP line.}
\end{minipage}
\end{figure}

\begin{figure}[h]
\begin{minipage}{36pc}
\includegraphics[width=36pc]{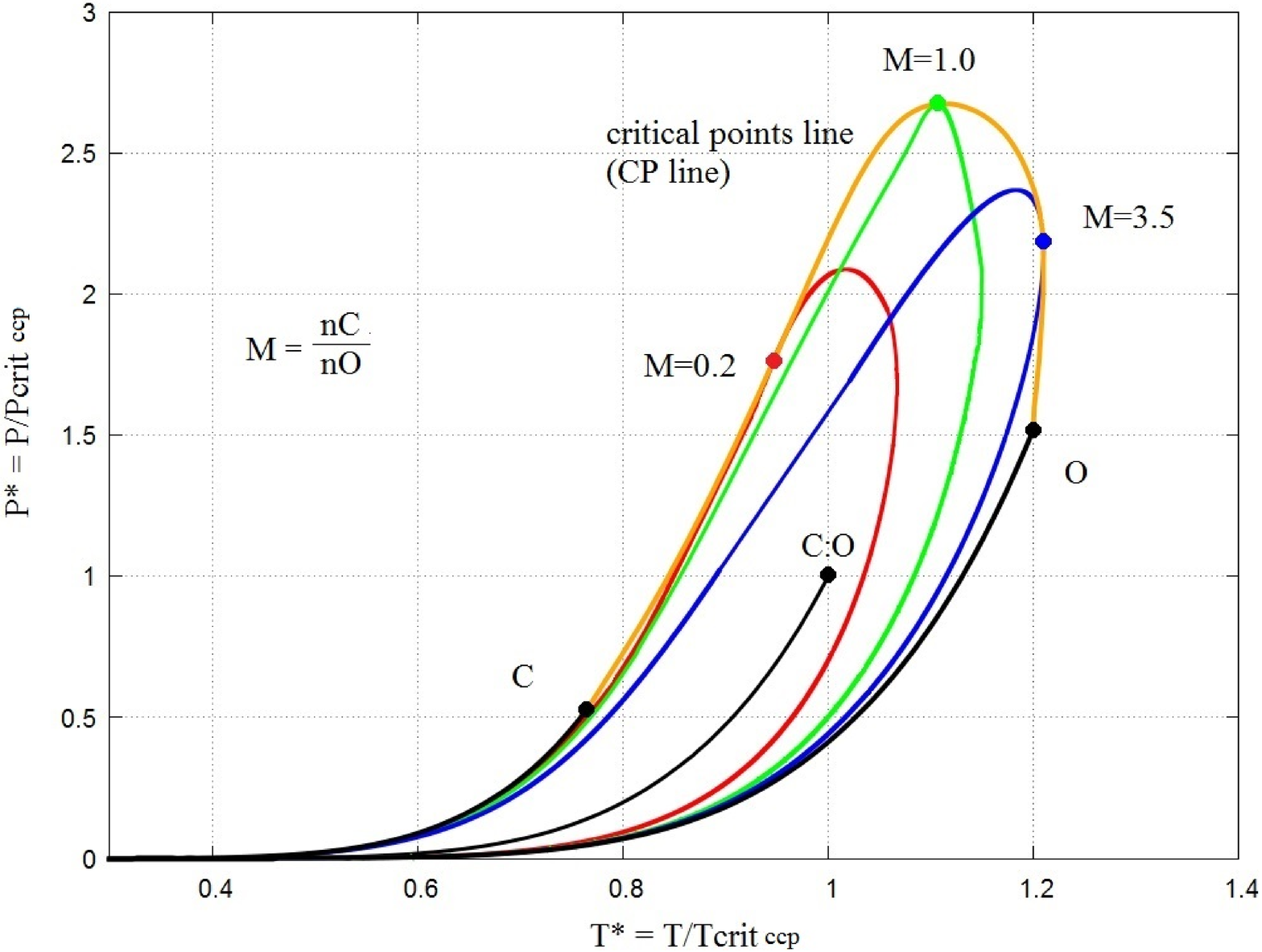}
\caption{\label{label5}P*-T* diagram for fully ionized plasma with different C-O composition. Black lines --- P*(T*) for pure C and O plasmas, and for C-O (1:1) mixture in forced-congruent regime. Red line --- P*(T*) boundary for two-phase region in non-congruent regime for C-O plasma enriched by carbon (M = 0.2), green --- the same for equimolar C-O mixture (1:1), blue --- the same for plasma enriched by oxygen (M = 3.5). Orange line --- the whole set of critical points (color online).}
\end{minipage}\hspace{1pc}%
\end{figure}

\section {The non-congruent regime of equilibrium (general case)}

	In general, the electrochemical potentials of the complexes,  of which the system consists, are not required to be equal, and according  to the Gibbs-Guggenheim equilibrium conditions, we obtain a system of equations for each component. In contrast to FCE the system of equations will contain $n-1$ more equations (where $n$ --- the  number of independent components), and hence more solutions. In this case, it leads to the existence of two solutions rather than one --- physically one of them corresponds to the beginning of boiling, the first appearance of vapor bubbles in the liquid phase with a given constant stoichiometry (further, BC --- boiling curve, a curve with a fixed ratio in the liquid phase, but with different gas stoichiometry), and the start of evaporation, the appearance of liquid droplets in a gas (further SC --- saturation curve, a curve with a fixed gas stoichiometry).

The two solutions are joined at a true critical point, that does not match with the CCP in FCE mode. Naturally the two-phase region occurs in $P-T$ coordinates, and the so-called "end" points --- with a maximum pressure (PMP --- pressure maximum point) and the maximum temperature (TMP), which in the same way do not necessarily match to the critical point (see figure 3). However, at a certain ratio of the components their pair  match can be observed (to reach a critical point of maximum pressure --- $M = 0.87$, temperature $M = 3.50$ with a certain accuracy in the study of carbon-oxygen mixture, where $M$ --- the ratio of the carbon ions concentration to the concentration of oxygen ions). There is also $\mu$\ MP (mu maximal point), which corresponds to the ration, when critical point math the maximum of electrochemical potential.

	The same ``banana-like" structure of 2-phase region was obtained in works \cite{1,9,10} while researching the non-congruence term in the uranium dioxide model. 2-phase region was also obtained in the study of liquid-gas phase transition in nuclear matter \cite{11}.

\section {The critical points line (CP line) in BIM($\sim$)}
Parameters of critical points line were calculated for the entire range of proportions of ions ($0<X<1$, where $X$ --- the ratio of carbon ions concentration to the total ions concentration, or $(1/1 + M))$, including the above characteristic values of CP coinciding with CP, PMP and TMP at $X_{T}= 0.222$, $ X_{P}=  0.535$ respectively, as well as the parameters of this line in congruent mode. There is a significant difference in the behavior of these curves, most notably manifested at an equal ratio of components (see figure 4). The position for such a two-phase region system also depends on the given stoichiometry. The figure shows the phase boundaries for different ratio of the components --- with a mixture of equal proportion, mixture depleted and enriched with carbon, respectively (see figure 5).

\section {The profile of electrochemical potentials and zeotropic BIM($\sim$)}
Another issue in the study of this model was to determine the azeotropic  properties.
A mixture of two or more components is azeotropic when its composition does not change during boiling, i.e. a mixture of compounds with the equilibrium equation of the liquid and vapor phases with the same stoichiometry. This means that there is a component ratio at which the transition from one state is possible without crossing the two-phase region, which implies that the hypothetical curves BC and SC must intersect at some point in the $P-T$ coordinates, and from the equilibrium conditions, it follows that the intersection will occur together with the FCE regime. Thus electrochemical potentials profiles component in $\mu$\-X coordinates should have the intersection, which are not observed in the calculations, then BIM is zeotropic mixture.

\begin{figure}[h]
\begin{minipage}{36pc}
\includegraphics[width=36pc]{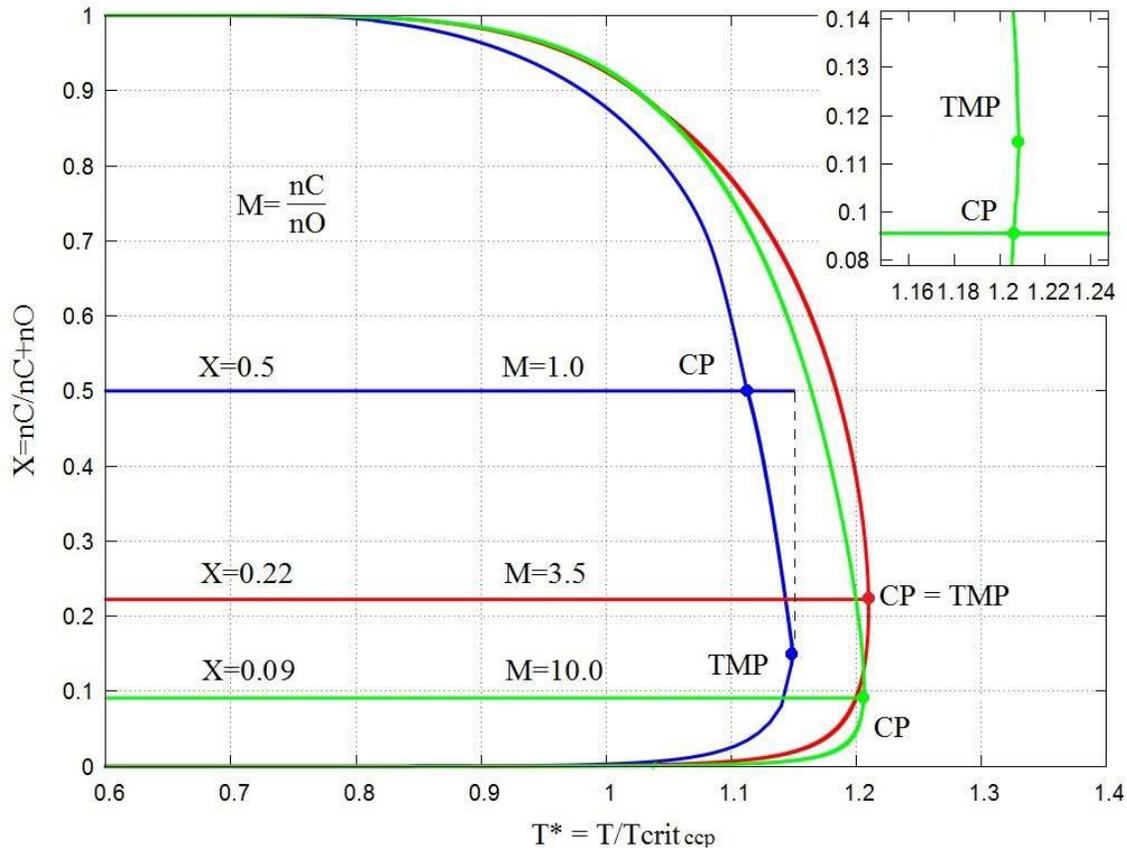}
\caption{\label{label6}X-T* diagram for mixtures with different stoichiometries in non-congruent equilibrium. The components proportions are given on the picture. The difference between TMP and CP is shown in the incut.}
\end{minipage}\hspace{1pc}%
\end{figure}

\section {$X-T$ diagram and the distillation effect}
	The stoichiometry of  one of the phases on BC and SC changes, as mentioned above (see figure 6). By lowering the temperature of the gas phase on BC and liquid phase on SC the distillation effect was observed --- the system advantageously consists of mainly the one component, while at the same time for comparison can be demonstrated in ~\cite{1,9,10} for chemically active plasma, where this effect is not present, and the phenomenon of incongruence drops at low temperatures. This distillation effect can be also observed in nuclear matter, which was mentioned above ~\cite{11}.

\begin{figure}[h]
\begin{minipage}{17pc}
\includegraphics[width=17pc]{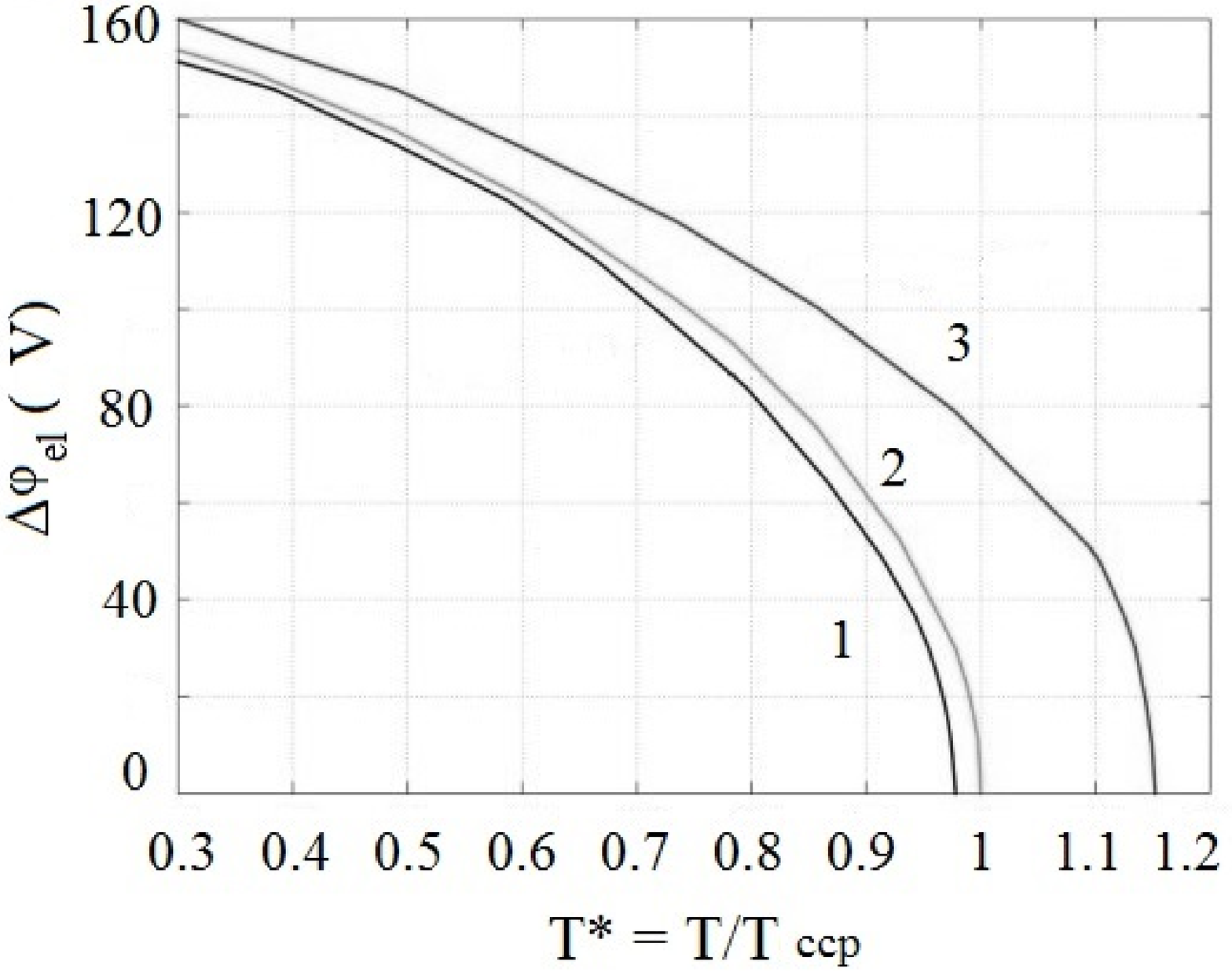}
\caption{\label{label7}d$\phi$(eV) -T*  diagram for the C-O mixture in two modes of equilibrium and nitrogen (1 --- nitrogen, as equimolar mixture, 2 --- FCE mode for C-O, 3 --- non-congruent mode for C-O).}
\end{minipage}\hspace{2pc}%
\begin{minipage}{17pc}
\includegraphics[width=17pc]{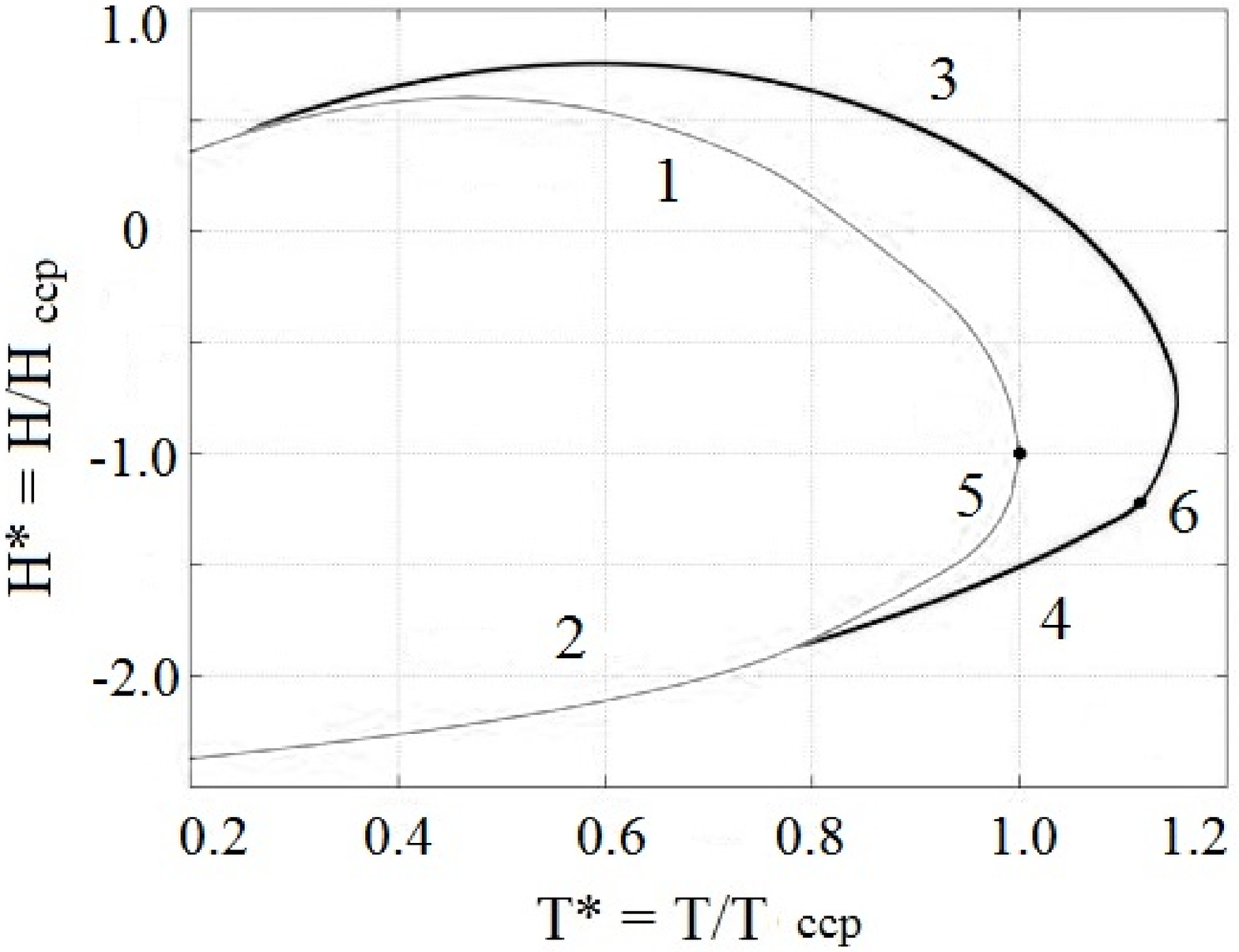}
\caption{\label{label8}H*-T*  caloric diagram for the C-O mixture in two modes of equilibrium with CP and CCP (1 --- congruent gas, 2 --- congruent liquid, 3 --- SC gas, 4 --- BC liquid, 5 ---congruent critical point, 6 --- critical point).}
\end{minipage} 
\end{figure}

\begin{figure}[h]
\includegraphics[width=17pc]{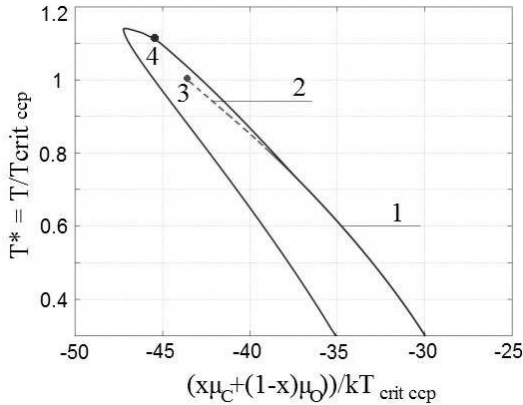}\hspace{2pc}%
\begin{minipage}[b]{17pc}\caption{\label{label}T*-$\mu$*  diagram for the C-O mixture in two modes of equilibrium with CP and CCP. Dotted line 2 --- FCE regime, bold 1 --- non-congruent mode, 3 --- CCP, 4 --- CP.}
\end{minipage}
\end{figure}

\section {The Galvani potential}
Electrostatics of interphase boundaries in Coulomb systems equilibrium (``Galvani potential") is a common case of the general laws for  such systems, which consists in the fact that any heterogenity in thermodynamic equilibrium Coulomb system is generally accompanied by the appearance of the electrostatic field, generated by the charges of the system itself\cite{3}.  The equilibrium potentials of electrons were explicitly taken from the calculated electrochemical potentials of carbon and oxygen (for carbon cell --- 6 electrons, oxygen, respectively 8, due to the condition of electroneutrality), which can not be equal. 
As in FCE and in  non-congruent mode of equilibrium the difference between these values in different phases is responsible for the above-mentioned Galvani potential.

This difference with increasing temperature decreases monotonically to the critical point, but the magnitude of the jump depends on the model and the regime of equilibrium (see figure 7).

\section*{References}
\bibliography{C:/Users/Nikita/Desktop/Articles/Elbrus/BibTeX/iopart-num/iopart-num.bib}

\end{document}